# XPS and DFT study of pulsed Bi-implantation of bulk and thin-films of ZnO – the role of oxygen imperfections

D.A. Zatsepin[1,2], D.W. Boukhvalov[3,4], N.V. Gavrilov[5], E.Z. Kurmaev[1,2], I. S. Zhidkov[2]

[1]*M.N. Miheev Institute of Metal Physics of Ural Branch of Russian Academy of Sciences, 620990 Yekaterinburg, Russia*
[2]*Institute of Physics and Technology, Ural Federal University, 620002 Yekaterinburg, Russia*
[3]*Department of Chemistry, Hanyang University, 17 Haengdang-dong, Seongdong-gu, Seoul 133-791, Korea*
[4]*Theoretical Physics and Applied Mathematics Department, Ural Federal University, Mira Street 19, 620002 Yekaterinburg, Russia*
[5]*Institute of Electrophysics, Russian Academy of Sciences, Ural Branch, 620990 Yekaterinburg, Russia*

*An atomic and electronic structure of the bulk and thin-film morphologies of ZnO were modified using pulsed Bi-ion implantation ($1 \times 10^{17}$ $cm^{-2}$ fluence, 70 min exposure under Bi-ion beam, $E_{Bi}^+$ = 30 keV, pulsed ion-current density of not more than 0.8 mA/cm$^2$ with a repetition rate of 12.5 Hz). The final samples were qualified by X-ray photoelectron core-level and valence band mapping spectroscopy applying ASTM materials science standard. The spectroscopy data obtained was discussed on the basis of DFT-models for Bi-embedding into ZnO host-matrices. It was established that in the case of direct Bi-impurities insertion into the employed ZnO-host for both studied morphologies neither the only "pure" $Bi_2O_3$-like phase nor the only "pure" Bi-metal will be preferable to appear as a secondary phase.  An unfavorability of the large cluster agglomeration of Bi-impurities in ZnO-hosts has been shown and an oxygen 2s electronic states pleomorphizm was surely established.*

E-mail: danil@hanyang.ac.kr

## 1. Introduction

The renewed research attention to zinc oxide has been emerged within last decades due to significant successful efforts in intensive development of sintering approaches. These approaches allow employing zinc oxide for fabricating ZnO-based electronic devices. Favorable aspects of employing ZnO as a prospective functional material include its radiation hardness, biocompatibility, thermal stability, optical, luminescence and electric properties (see i.e. Refs. [1-3]). Having several known phases – hexagonal wurzite, cubic zincblende, hi-pressure rocksalt (NaCl) and hi-temperature cesium chloride [4] – zinc oxide demonstrates rather wide range of interconnecting types for $Zn^{2+}$ and $O^{2-}$ sublattices and degrees of electron sharing. On the one hand, this results in experimentally detected Zn–O and O–Zn–O bonding deviations from the commonly known ideal unit-cell parameters for the concrete arrangement type of crystalline structure (more detailed described in Ref.[5]), but, on the other, zinc is always divalent in all these zinc-oxygen structural phases. It is believed, that mentioned features might be responsible for the attractive chemical durability of ZnO while employing it as a functional base for the photocatalysis applications [6] and further band-gap engineering. At the same time well known, that the concentration of native structural defects in ZnO (agglomerates of oxygen vacancies and/or zinc interstitials), which were found to be responsible at least for visible photoluminescence [7], is strongly affected by the initial preparation method as well as by the posterior treatment. Moreover, these defects strongly manipulate in a negative way the photostimulated conduction, electric and catalysis frontier properties of ZnO-based materials. This means that a proper choice in sintering approach from the very beginning is the fundamental basis for a successful engineering and functional modifications of the electronic structure in the ZnO-based electronic devices.

An undoped as-sintered ZnO is usually of *n*-type conductivity, whereas doping with Group V elements (such as As, Sb, Bi) is of great importance in terms of attempt to switch on the *p*-type. At least, previously made doping ZnO with As at concentrations of $10^{19}$ $cm^{-3}$ shows the SIMS results which are quite close to *p*-type nature [8]. Unfortunately the secondary phase side-effect was found (in this concrete example – $Zn_3As_2$) which have to be considered while selecting the fabrication method for ZnO:Me (Me = As, Sb, ...) system. Recent results [9, 10] also demonstrate that implantation of ions of various metals under used sintering regimes provides the formation of the Zn-site substitution or/and Zn-site vacancies. The latter itself plays an important but negative role for the high *p*-type conductivity because can easily trap charge-carriers. Nevertheless, these vacancies (the former Zn-site positions) are quite useful for injected Me-particles in order to be occupied by them. More detailed the effect of vacancies on *p*-conductivity is discussed in Refs.[7, 11-12].

As for Bi-ion implantation of ZnO and modification of its electronic structure, there is, unfortunately, a lack of published results so this question needs further study. It is clear from general considerations that the stumbling block can become an essential ionic radii mismatch for Zn and Bi while implanting ZnO host with Bi-ions. This might cause the strongest damage of as-sintered initial ZnO host-structure and, probably, an overall Zn–O and O–Zn–O bonding nature re-arrangements. The most interesting seems the final structure of Bi-implanted zinc oxide in terms of physical chemistry key-point – what will be Zn and Bi final valence states after ion-beam treatment. This rhetorical question stems from the fact that both metals are mostly retaining the valency in their known variety of Zn–O and Bi–O structural polymorphs. From the aforesaid we are formulating the main idea of our research, the results of which will be presented in the current paper: the direct XPS and DFT "as is" structural transformations of Bi-ion implanted ZnO-host produced by pulsed mode of Bi-embedding. Similarly to our previous works [13-15] we will discuss the difference between Bi-implantation of the bulk and surface (thin-film) morphologies on the basis of almost unavoidable oxygen vacancies. This issue is important because an atomic structure of semiconducting oxides, such as $TiO_2$ and ZnO, usually contains enough space for appearance of interstitial vacancies and formation of clusters of substitutional and/or interstitial defects, so this have to be taken into consideration while employing th model calculations. The most attention will be paid to the difference in scenarios of Bi-impurities distributions between ZnO and $TiO_2$. In contrast with titanium dioxide, the mismatch between oxidation states of "removed" host-ions and impurity-ions as well as the dissimilarity in native lattices will play a crucial role in ZnO:Bi. The results obtained in this work should be the next step toward understanding general principles of impurities incorporation and distribution in semiconducting oxides.

## 2. Samples, Experimental and Computational Details

ZnO *bulk ceramics* samples were made employing hot-compacting procedure (see, for instance Ref. [16]) of the appropriate ceramic powders in the oxygen ambient with posterior tempering at 1000 °C. This yields the samples with 5.6 g/cm$^3$ density and standard average dimensions of 13 mm in diameter and 2 mm in height (mostly suitable size for mounting into the sample-holders for various experimental systems). The fabricated samples were qualified by X-ray diffraction (XRD) measurements which display that *bulk ceramics* is in zincite single phase state with a hexagonal structure ($a$ = 3.251 Å, $c$ = 5.202 Å) and average crystallite size of about 200 nm. To grow the ZnO *thin-films*, a sapphire substrate (100) was used for their deposition which had been ultrasonically cleaned in acetone and alcohol for 10 min, then rinsed in deionized water, and finally dried. The growth rate of ZnO *thin-films* was 3.4 nm/min and the typical thin-film thickness was not more than 30 nm. The yielded *thin-film* ZnO samples had a hexagonal structure with lattice parameters $a$ = 3.250 Å and $c$ = 5.207 Å. For deeper details of sintering refer to Ref. [17].

Bi-ion implantation of the *bulk* and *thin-film* ZnO samples was carried out in a pulsed-repetitive mode at the residual gas pressure of $1 \times 10^{-3}$ Pa with MEVVA-type ion source (see, for example Ref. [14]). Bi-ion fluence (integrated flux) of $1 \times 10^{17}$ cm$^{-2}$ was achieved after 70 min of ion exposure. Bi-ion beam with ion energy of 30 keV, pulsed current density of not more than 0.8 mA/cm$^2$ and repetition rate of 12.5 Hz with pulse duration of 0.3 ms were applied for Bi-embedding. The average temperature of the samples during ion implantation did not exceed 295 °C in this case. No posterior tempering was made for both morphologies of the samples under study.

The chemical purity of initial (as sintered) and Bi-implanted ZnO samples was additionally qualified with the help of PHI XPS Versaprobe 500 spectrometer (ULVAC–Physical Electronics, USA), applying Al X-ray $K\alpha$ radiation (1486.6 eV) [15]. The same XPS system had been employed for posterior core-level and valence band mapping analysis of the samples under study with the help of XPS Databases [18-20] using a mutual cross-check in. The results of XPS qualification (fast wide-scan or survey spectra) of Bi-implanted zinc oxide host-matrices in the form of *bulk* and *thin-film* morphologies versus reference ZnO XPS external standard are presented at Fig.1.

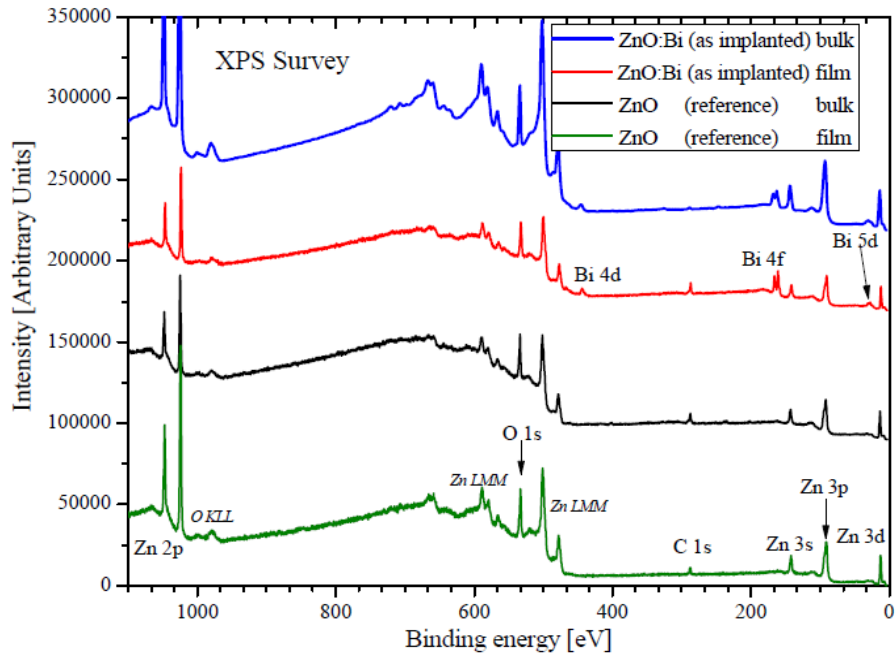

**Figure 1.** XPS Survey spectra for ZnO:Bi in the *bulk* and *thin-film* morphologies without tempering. Also reference survey spectra for as sintered ZnO in the appropriate morphologies are shown for comparance.

The detected and identified photoelectron peaks allow to conclude that there are no alien impurities in implanted *bulk* and *thin-film* ZnO hosts except those belonging to embedded Bi-implant. Carbon 1*s* signal at 285.0 eV is relatively low comparing with the other XPS peaks but, at the same time, is well enough in order to perform precise calibration using ASTM Standard Guide [21]. No extra contaminations within the sensitivity range of applied XPS method were found and the declared empirical formulas for the samples under study are valid. The valence density of states BE region together with core-levels and core-like bands, ranging from 0 eV up to ~ 30 eV, will be analyzed onwards by means of XPS Valence Band Mapping approach.

The density-functional theory (DFT) calculations were performed using the SIESTA pseudopotential code [22] as had been previously used for related studies of impurities in the bulk and thin-film semiconductors [13-15]. All calculations were made using the Perdew-Burke-Ernzerhof variant of the generalized gradient approximation (GGA-PBE) [23] for the exchange-correlation potential. A full optimization of the atomic positions had been applied, during which the electronic ground state was consistently found using norm-conserving pseudopotentials [24] for the cores and a double-$\xi$ plus polarization basis of localized orbitals for Bi, Zn, and O. The forces and total energies were optimized with an accuracy of 0.04 eV/Å and 1.0 meV, respectively. The calculations of the formation energies ($E_{form}$) were performed by considering the supercell both with and without a given defect [15]. The supercells consisting of 108 atoms were used as a host for studying defects in ZnO (see Fig. 2).

Taking into account our previous modelling of transition metal impurities in semiconductors [13-15], we have calculated various combinations of structural defects including single substitutional (1S) Bi-impurity, pairs of substitutional impurities (2S, Fig. 2a) and their combinations with interstitial (I) impurities (S + I and 2S + I) and oxygen vacancies (+vO, see Fig. 2b-d). For the modeling of ZnO surface we applied the same supercell as a slab (Fig. 2c,d) which is feasible model of (001) surface of zink oxide [15].

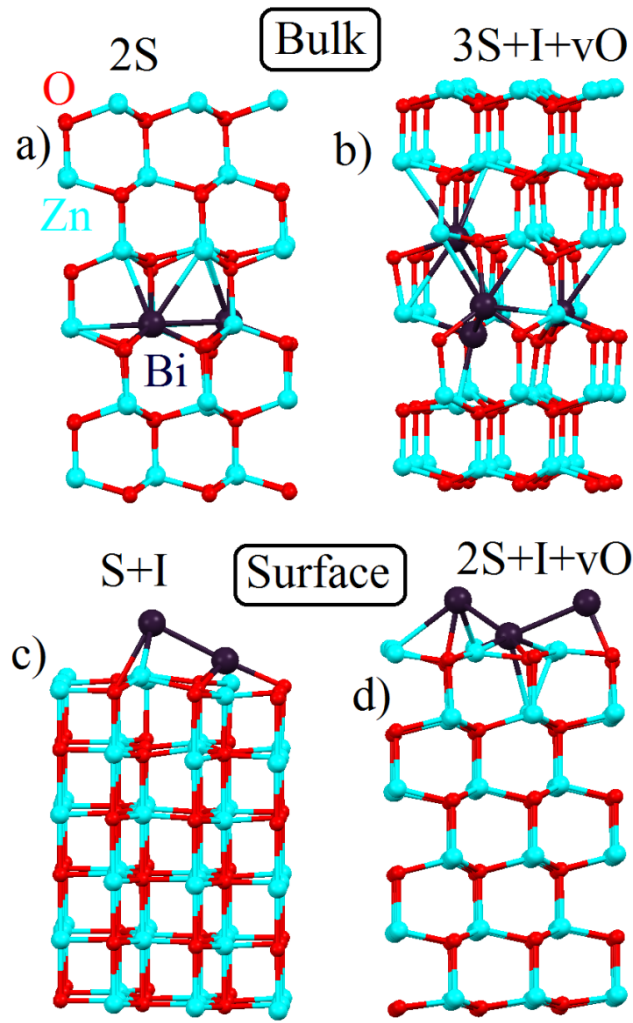

**Figure 2.** An optimized atomic structure of various combinations of substitutional (S) and interstitial (I) bismuth defects and oxygen vacancies (vO) in bulk (a,b) and on surface (c,d) of ZnO.

## 3. Results and Discussion

Since Bi was implanted into "metal-oxygen" matrix, there is a high probability of Bi solid-state interaction with an oxygen sublattice of employed target-host, leastwise due to ballistic interatomic collisions. With this connection, it seems logically justified to make comparisons among the appropriate XPS spectra of our samples, native metal and stable metal oxide phases for implanted Bi. This will allow defining clearly whether the implanted Bi remains in a metal state as a metal clusters or forms one of the known bismuth-oxygen pleomorphs. XPS Bi 4f core-level spectra for Bi-implanted zinc oxide host-matrices in different morphologies and appropriate

XPS external standards (Bi-metal and α-Bi$_2$O$_3$) are shown at Fig. 3. The recorded from ZnO:Bi samples the core-level 4f XPS spectra have a complex structure, which demonstrate the presence of low-intensity sub-bands denoted at Fig.3 as *B*-type features and high-intensity *A*-type features. Their detailed spectral parameters are presented in Table I for clarity.

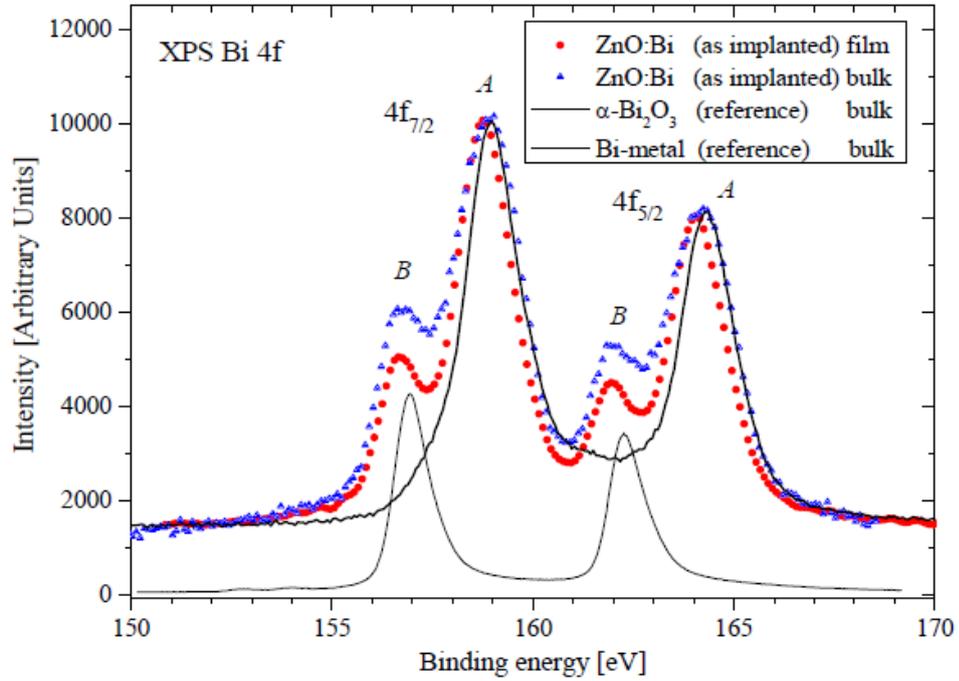

**Figure 3.** XPS Bi 4f core-level spectra for ZnO:Bi in the *bulk* and *thin-film* morphologies without tempering and reference spectra for α-Bi$_2$O$_3$ and Bi-metal (XPS external standards).

**Table I.** XPS parameters Bi 4f core-level spectra for ZnO:Bi in different morphologies and XPS external standards.

| Sample or XPS standard (reference) | Core-level component Binding Energy position, eV | | Spin-orbital separation Δ, eV |
|---|---|---|---|
| | Bi 4 f$_{7/2}$ | Bi 4 f$_{5/2}$ | |
| α-Bi$_2$O$_3$ (reference) | 158.97 (*A*) | 164.30 (*A*) | 5.33 |
| ZnO:Bi (bulk) | 158.89 (*A*) | 164.22 (*A*) | 5.33 |
| ZnO:Bi (film) | 158.84 (*A*) | 164.16 (*A*) | 5.32 |
| Bi-metal (reference) | 156.93 (*B*) | 162.26 (*B*) | 5.33 |
| ZnO:Bi (bulk) | 156.82 (*B*) | 162.15 (*B*) | 5.33 |
| ZnO:Bi (film) | 156.78 (*B*) | 162.11 (*B*) | 5.33 |

From Figure 3 and Table I becomes clear, that 4f $_{7/2}$ and 4f$_{5/2}$ spectral bands of *A*-type are almost identical to that for α-Bi$_2$O$_3$ which was applied as a reference, whereas the *B*-type features are much closer to that for Bi-metal XPS reference. As an additional testimony the nearly symmetrical spectral shape for *A*-type of Bi 4f components in ZnO:Bi spectra might be considered (it is usually specific for bismuth oxide XPS Bi 4f spectra [18]). In this way we have all the reasons to interpret the described 4f components of *A*-type in both morphologies of ZnO:Bi as an indication of occurred bismuth-oxygen interactions during Bi-implantation. As it is well known, XPS core-level spectra of "pure" (not even slightly oxidized) metal have strongly asymmetrical line-shapes [25], that is an additional XPS signature for an identification of the presence of "pure" metal phase in the sample under study. One can see from Figure 3, that all *B*-type features also have a very noticeable line-shape asymmetry. Thus in terms of this, we might suppose that the embedded into zinc oxide Bi-metal oxidizes only partially and some of metal particles are probably translocated into interstitial positions of zinc-oxygen host-structure, giving the rise of the low-intensity XPS *B*-type Bi 4f components. Quite possible and sane reason for that is the great mismatch of ionic radii for Bi and Zn in the applied for implantation ZnO host-structure. This situation will be modeled onwards in the current paper. Finally, the formal valence states of implanted Bi on the basis of XPS core-level analysis were detected as Bi$^{3+}$ and Bi$^0$.

XPS Zn 2p core-level spectra of as implanted ZnO:Bi in the *bulk* and *thin-film* morphologies are shown at Fig.4. These spectra have a simple double-peak structure due to Zn 2p$_{3/2}$ − 2p$_{1/2}$ spin-orbital separation and typical line-shapes that were established for zinc oxide [18-20]. As it was reported in Ref. [18], the Binding Energy (BE) for Zn 2p in as-grown ZnO is nearly 1022 eV (Zn 2p$_{3/2}$) and for Zn-metal is 1021.7 eV (Zn 2p$_{3/2}$), respectively.

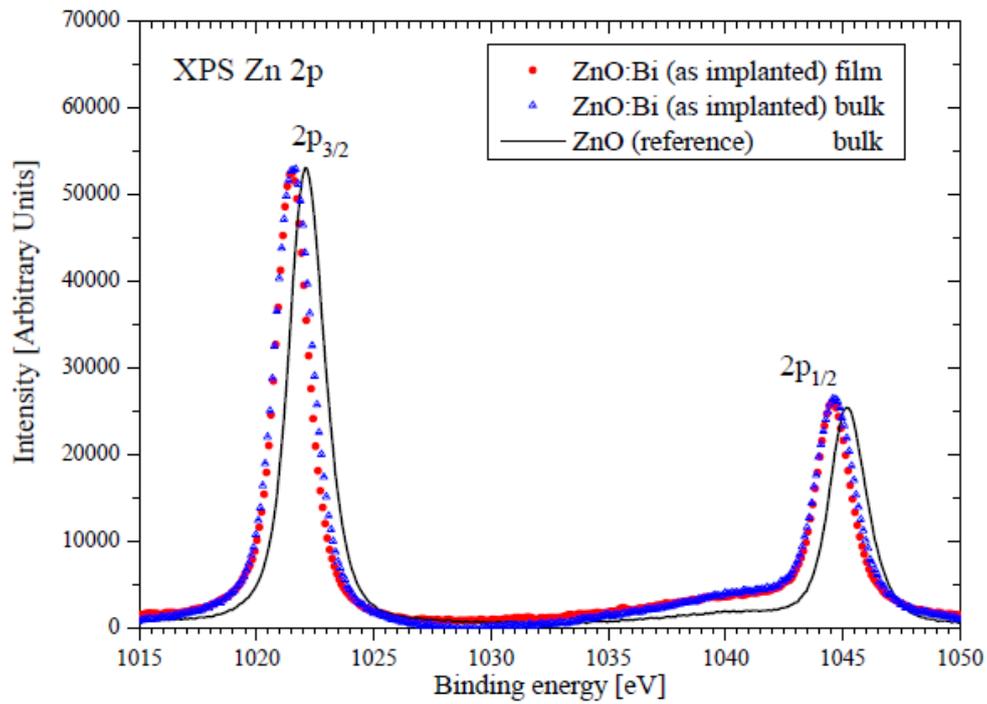

**Figure 4.** Comparance of XPS Zn 2p core-level spectra for ZnO:Bi in both studied morphologies.

This very small BE-shift, while varying the valence state from $Zn^0$ to $Zn^{2+}$, is impeding the chemical state differentiation of Zn in zinc-containing compounds, because if an XPS system has unsatisfactory energy resolution and signal-to-noise ratio, there will be no chance for detecting the actual values of this BE-shift. Nevertheless we were able to resolve this chemical differentiation and our values of BE's are the following for Zn $2p_{3/2}$ component: 1021.56 eV (ZnO:Bi bulk) and 1021.57 eV (ZnO:Bi thin-film), i.e. these are nearly the same as in Zn-metal [18]. Also the Zn $2p_{3/2} - 2p_{1/2}$ line-shapes has some asymmetry while comparing with that for ZnO (see Fig.4). So we might suppose that Bi-implantation of ZnO on the ballistic stage of this process breaks Zn–O bonding and produces some amount of Zn-vacancies in the structure of ZnO:Bi. Since the XPS Zn 2p core-level spectra are identical in both morphologies of ZnO:Bi, one might suppose that the process of Zn-vacancies formation occurs in them in a similar manner.

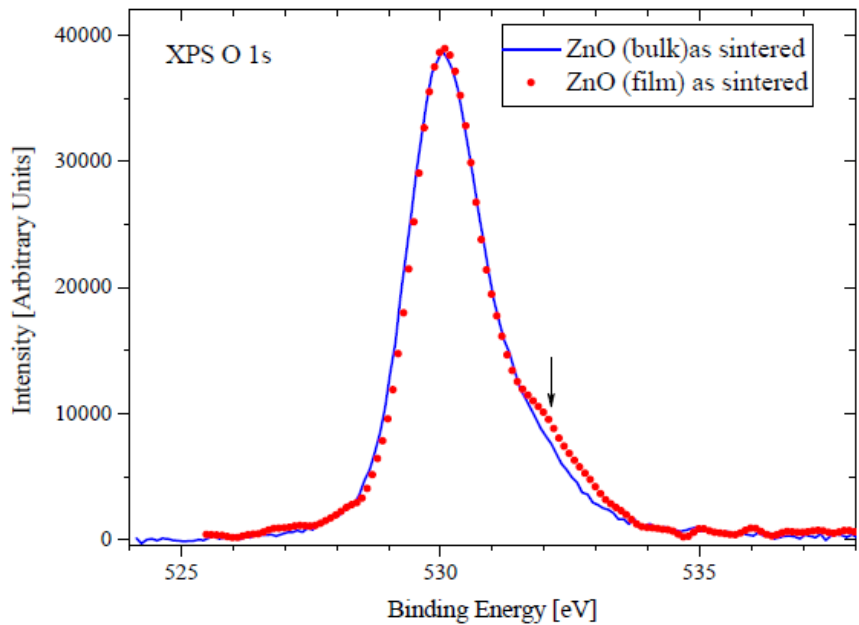

**Figure 5.** XPS O 1s core-level spectra of as sintered ZnO host-matrices in both morphologies.

The question of natural oxygen vacancies in ZnO atomic structure needs some brief discussion from materials science point of charge. The first valuable and precise XPS data concerning the nature of O 1s core-level line-shape in zinc oxide was reported in Refs. [25-27] and even included into international NIST XPS Database [18]. The reason for that is linked with the relatively complex line-shape of XPS O 1s core-level in ZnO because of the hi-sensitivity to oxygen sublattice imperfections, which are arising in XPS O 1s spectrum as a wide low-intensity sub-band at ~ 532 eV (see Fig. 5). At the same time, Deroubaix *et. al.* [27] were discussing the origin of the very similar sub-band (they obtained the BE value of 532.27 eV) as a contribution of OH-group, but they were using $Zn(OH)_2$ as a material under study. They also mentioned that such contribution also might be spectrally visible in a "dry" ZnO due to some presence of hydroxyls. Nevertheless, in Ref. [28] a detailed XPS testimony was presented about the oxygen imperfection nature of oxygen vacancies type for ~ 532 eV sub-band. The same point of view was independently and reasonably reported by the authors in Refs. [29-31]. If we will take into account that no "wet" technology was applied for preparation of our ZnO host-matrices (viz. no OH-groups are present in our samples – see section *Samples, Experimental and Computational*

*Details*), so the origin of ~ 532 eV sub-band well coincides with the interpretation as an oxygen sublattice imperfections of a vacancy type. Moreover, an OH contribution to O 1s core-level of ZnO is positioned at 533.1 eV, having specific and dissimilar line-shape [30]. In terms of such an interpretation, both types of ZnO host-matrices after Bi-ions implantation have similar configuration of O 1s core-levels with slightly higher imperfections of oxygen sublattice in the *thin-film* morphology comparing with the bulk one (Fig. 5). As for the main high-intensity XPS peak in O 1s core-level spectrum, it is placed at 530.05 eV with the line-shape well coinciding with the data reported in Refs. [19, 25, 28-31]. Having the materials science basis for the onward interpretation of our XPS data, we will now identify the O 1s core-level spectra for ZnO:Bi in the *bulk* and *thin-film* morphology. These spectra are shown at Fig. 6.

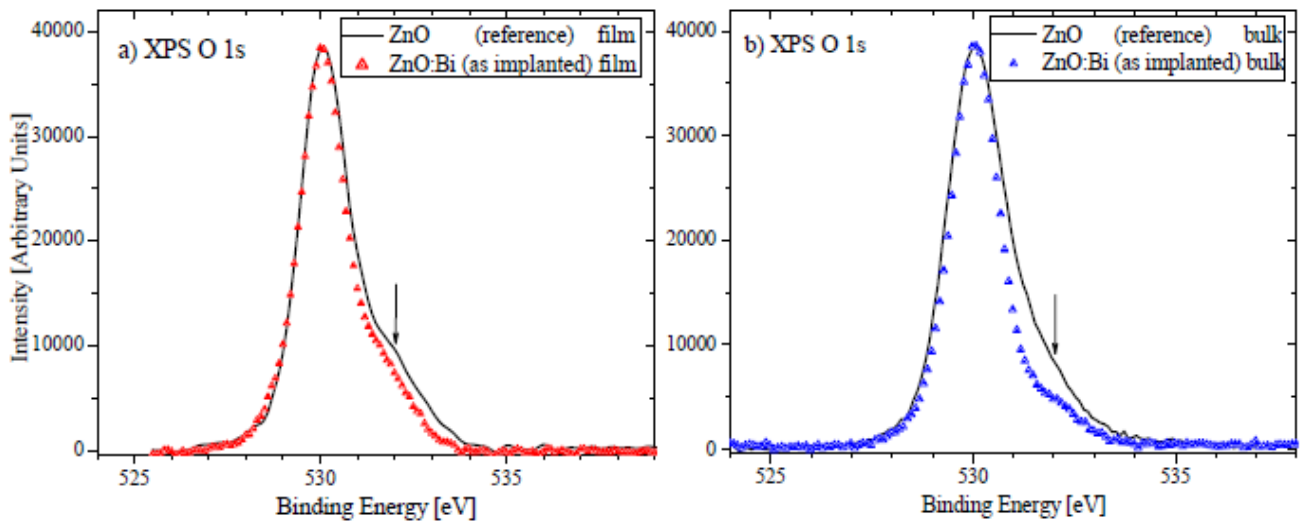

**Figure 6.** XPS O 1s core-level spectra of ZnO:Bi in the *thin-film* (a) and *bulk* (b) morphologies.

For *thin-film* of ZnO:Bi one can see the only weak difference in O 1s core-level (Fig. 6a), whereas for the *bulk* morphology the dissimilarity in 532 eV sub-band is much notable (Fig. 6b). Basing on the above discussed interpretation, the mentioned line-shape and intensity differences between reference (as sintered) ZnO and ZnO:Bi might be a sign of different imperfections in Zn–O bonding after Bi-implantation for the *bulk* and *thin-film* ZnO host-matrices. Namely, opposite to the *bulk* the *thin-film* morphology does not transforming radically its oxygen sublattice arrangement type as well as imperfections concentration. At least the very close effect was found in ZnO doped with Hf, where Hf precipitation reduced oxygen vacancies concentration in ZnO and enhanced Zn–O bonding strength (vacancy treatment effect) [32]. Our XPS data allows to state that we have gotten the very close effect but induced by pulsed Bi-implantation. The reported data seems to be a point factor for the link of the cluster structure stability with the *thin-film* ("surface") and the *bulk* morphologies of the newly composed ZnO-based functional material. This will be checked with the onward DFT-modeling approach of an electronic structure of the ZnO:Bi oxides.

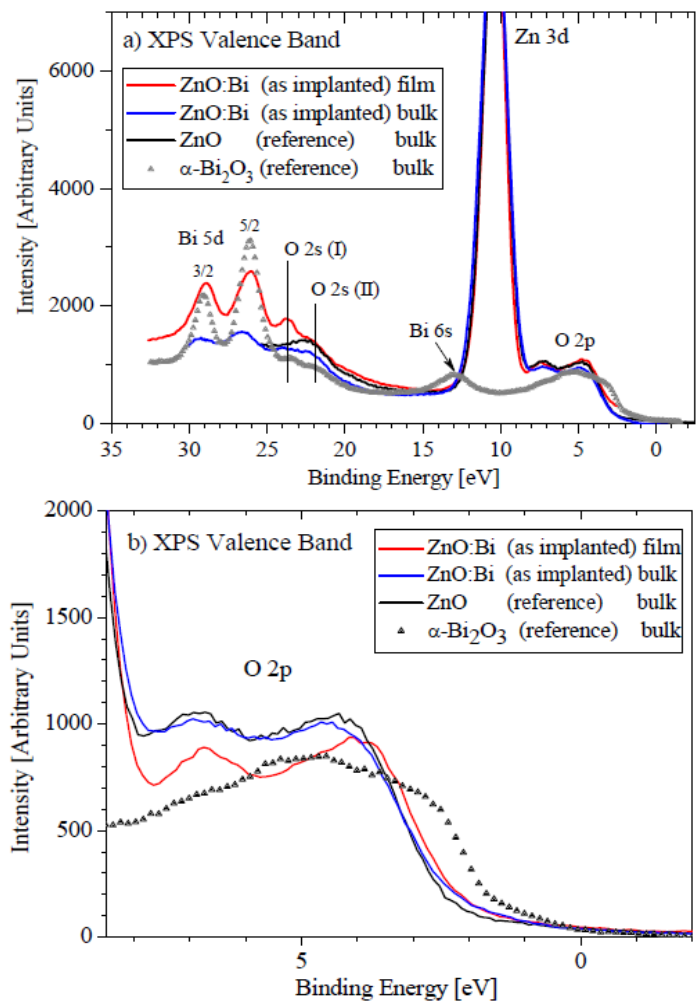

**Figure 7.** (a) XPS Valence Band mapping for ZnO:Bi in the *thin-film* and *bulk* morphologies; (b) Zoomed XPS Valence Band region for the same samples.

The results of XPS Valence Band (VB) mapping for the samples under study are presented at Fig. 7. One can clearly see that the VB of reference α-$Bi_2O_3$ has true to type Bi $5d_{5/2-3/2}$ core-like bands located at ~ 25.4 eV and 28.3 eV, respectively. These Bi 5d bands are absent in VB of reference ZnO spectrum, which is not surprising, but clearly detected in the valence band spectra of Bi-implanted ZnO *thin-films* and *bulk* (see Fig. 7a), indicating, at least, the presence of embedded bismuth 5d states in Bi-implanted samples. The valence band range from about 21eV up to 24 eV usually belongs to oxygen 2s states in oxide materials [18-20]. Due to pleomorphic origin of chemical Bi–O bonding in α-$Bi_2O_3$ [33,34], the valence band of this oxide exhibits the distinct and specific contribution of O (I) and O (II) 2s partial electronic states to VB in this BE-region, that is surely characteristic "fingerprint" for detecting whether embedded Bi is in oxide form or in the form of Bi-metal clusters. Our previous combined XPS and DFT study of $TiO_2$:Bi [32] well agrees with the point of view reported in Refs. [30-31] and displays the same separation between pleomorphic O (I) and O (II) 2s electronic states for α-$Bi_2O_3$ oxide and Bi-implantation sintered α-$Bi_2O_3$ phase in employed $TiO_2$ host-matrix. These distinct O (I-II) 2s XPS bands are also seen in the XPS VB spectra of ZnO:Bi in the *thin-film* and *bulk* morphologies (Fig. 7a), clearly indicating an oxidation of implanted bismuth particles. The intensity of O (I-II) 2s XPS bands for the *bulk* morphology of ZnO:Bi is significantly lower than that for *thin-films*.

At the same time, in the XPS Bi 4f core-level spectra for the *bulk* ZnO:Bi one can see much stronger XPS signal form bismuth in metallic phase (*B*-band) in comparance with *thin-films* (see Fig. 3). Also the ~ 532 eV band in XPS O 1s core-level spectrum, which is linked with oxygen imperfections, as it have been discussed above, is much lower than that for implanted film of ZnO (Fig. 6a-b). So in summary we might conclude that in the *bulk* morphology of ZnO the implanted Bi is predominantly accumulated in a metal-cluster form rather than oxide, but for the *films* an opposite situation occurs. As for vicinity electronic states (Fig. 7b) we have not detected any essential transformation of electronic structure of employed ZnO host-matrices opposite to the case of TiO$_2$ [13]. One of the possible reasons for such unusual behavior might be the suppressing of Bi-contribution to vicinity region of VB by Zn 3d partial electronic states which are exhibiting an extremely high-intensive XPS signal (see Fig. 7a), and the majority of dominating the O 2p partial electronic states in the range of ~ 3 eV – 6.5 eV. The latter is obviously usual for the valence band structure of oxide materials and not so surprising. The high-intensive XPS Zn 3d signal also might be a sign of Zn-metal clustering in the interstitial positions. Recall, that XPS Zn 2p core-level spectra has the very close parameters to that for Zn-metal (see Fig. 4).

For the onward and extended discussion of XPS data obtained we applied the formalism of DFT-theory for the electronic structure calculations and evaluate energetics of various defects in pure ZnO and in the presence unavoidable oxygen vacancies. The results of calculations (see Table II) are demonstrating that the absence of oxygen vacancies in the *bulk* ZnO is admitting the formation of substitutional Bi-defects (1S and 2S) and also possibility of moving Zn-ion of host matrix into the interstitial void. The presence of oxygen vacancies dramatically changes the scenario. Similarly to the insertion case of large Bi-impurities in TiO$_2$ host-matrix [13-15], oxygen vacancies create an additional extra empty space which is decreasing the values of formation energies for Bi-clusters in the form of substitutional and interstitial impurities. In order to check the transition from the initial stages of fabrication of these clusters (S+I) to the stage of separate bismuth-metal phase formation, we performed the calculations for the various configurations, viz. nS+I (+vO) with n ranging from 1 to 6. It was established (see Table II) that the lowest formation energies correspond to the smallest-size clusters and the 4S+I+vO configuration starts to grow up. This is crucial difference between TiO$_2$:Bi [13] (where oxygen vacancies provide the clustering of metallic Bi-phase) and ZnO:Bi, caused by more dense package of ZnO lattice. In contrast with nS+I clusters of 3d-metals in ZnO [15], 2S+I and 3S+I clusters does not provide any valuable change in electronic structure of ZnO (Fig. 8a) similarly to the substitutional impurities. In both cases only a small contribution from Bi 6s states appears and in the case of 3S+I clusters formation the separation of O 2s band into distinct but very close in energy O(I) and (II) peaks occurs (oxygen pleomorphizm). This is in a good agreement with an experimental XPS data presented at Fig. 7. The presence of metallic bismuth *B*-features in XPS core-level spectra (see Fig. 3) might be caused by the appearance of Bi–Bi bonds in ZnO:Bi. These bonds could appear only in the case of nS+I Bi-clusters which could be formed and located only in the vicinity of oxygen vacancies in ZnO, having essentially limited size (e.g. below four bismuth atoms, see Table II). If so, they does not provides any valuable changes in VB and that is seen in experimental XPS Valence Band spectra (compare Fig's. 7 and 8).

**Table II.** Calculated formation energies (eV per defect atom) for various defects in the *bulk* (Fig. 2a,b) and "surface" (*film*) (Fig.2c,d) ZnO without and with oxygen vacancies (vO) in vicinity of defects. The most probable defects are marked with bold font.

| Defect | Bulk | | "Surface" (film) | |
|---|---|---|---|---|
| | pure | +vO | pure | +vO |
| 1S | **2.09** | 4.13 | **-3.78** | 1.64 |
| 1S+vZn | 3.41 | 3.66 | 1.84 | 2.43 |
| 1S+ZnI | **2.86** | 3.12 | 0.98 | 1.51 |
| 2S | **2.66** | **3.05** | -2.29 | **0.33** |
| S+I | 4.46 | **3.08** | -2.34 | **-0.41** |
| 2S+I | 4.21 | **2.91** | -1.53 | 0.64 |
| 3S+I | 3.36 | **2.82** | 0.21 | 1.28 |
| 4S+I | 3.44 | 3.44 | 1.45 | 1.78 |
| 5S+I | 3.58 | 3,55 | 2.69 | 3.00 |
| 6S+I | 3.83 | 3.61 | 2.82 | 3.07 |

Similarly to the insertion case of Bi-impurities into the *bulk*, the substitutional impurities on the surface are the most favorable type of defects. But in contrast with the *bulk* morphology the formation of smallest S+I clusters might also occur without any presence of oxygen vacancies. The presence of oxygen vacancies makes the formation of substitutional defects less favorable than S+I pair. The formation of this pair provides the base for the appearance of Bi–Bi bonds which are responsible for Bi-metal *B*-features in XPS Bi 4f core-level spectra (Fig. 3) on the one hand, and, on the other, also results in O 2s band separation into pleomorphic sub-bands (see Figs. 7a and 8b). All these performing the visible shift of highest occupied energy level of the valence band (Figs. 7b and 8b). In contrast with 3d-impurities insertion case, the "surface" of ZnO simply accumulating Bi-atoms in the form of nS+I clusters which provides significant shift of Bi-atoms above "surface" (Fig. 2d). As a reply, one can see the notable increasing in formation energy (Table II), so the formation of clusters like 2S+I and appearance of metallic features in valence band (Fig. 8b) becomes essentially unfavorable.

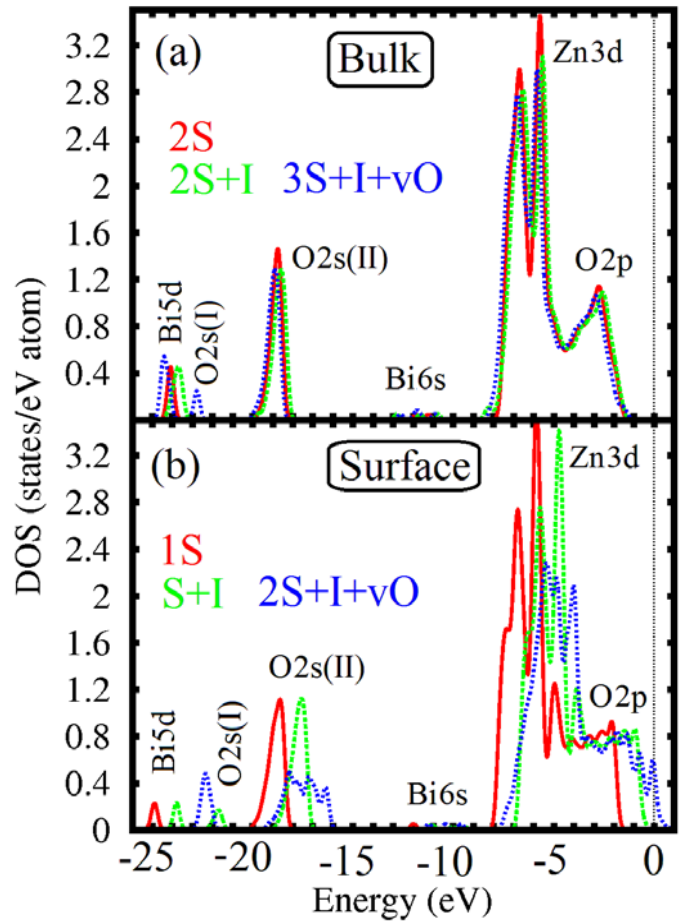

**Figure 8.** Total densities of states (DOS) for various configurations of substitutional (S) and interstitial (I) bismuth defects in the presence of oxygen vacancies (vO) for the *bulk* (a) and "surface" (b) morphologies of ZnO-host.

## 4. Conclusions

Based on the results of XPS core-level and XPS valence band mapping analysis and DFT-modeling we have demonstrated that in the case of direct Bi-impurities insertion into the employed ZnO-host for both studied morphologies neither the only "pure" $Bi_2O_3$-like phase nor the only "pure" Bi-metal will be preferable to appear as a secondary phase. For the both employed type of morphologies (*bulk* and *thin-films*) the single substitutional defects are energetically preferable. The superposition of Bi-metal features with bismuth oxide features in XPS Bi 4f core-level spectra in the *bulk* and *thin-films* in combination with the absence of any visible VB spectra re-arrangement in the vicinity of Fermi level might be caused by the fabrication of small-size clusters of a few substitutional and interstitial Bi-impurities, located as nearest neighboring for oxygen vacancies in ZnO-host. In these small-size clusters the presence of Bi–Bi bonds are the reason for appearance of metal-like *B*-features in the XPS Bi 4f core-level spectra but they are not transforming the electronic structure of valence bands significantly at least within the range of sensitivity of XPS method. Our theoretical considerations, based on XPS data reported; also demonstrate an unfavorability of the large cluster agglomeration of Bi-impurities in ZnO-hosts. At the same time an oxygen pleomorphizm (as in the case of $TiO_2$:Bi) was surely established.

## Acknowledgements

The synthesis of ZnO samples and posterior Bi ion-implantation treatment were supported by the Act 211 of the Government of the Russian Federation (Agreement No. 02.A03.21.0006) and the Government Assignment of Russian Ministry of Education and Science (Contract No. 3.1016.2014/K). The XPS qualification and analysis of the samples under study were supported by the Russian Science Foundation (Project No. 14-22-00004).